\documentclass{article}
\usepackage{arxiv} 
\usepackage{authblk}
\usepackage{graphicx}

\usepackage{comment}
\usepackage{amsmath}
\usepackage{hyperref}
\usepackage{xcolor} 

\usepackage{listings}
\definecolor{codegreen}{rgb}{0,0.6,0}
\definecolor{codegray}{rgb}{0.5,0.5,0.5}
\definecolor{codepurple}{rgb}{0.58,0,0.82}
\definecolor{backcolour}{rgb}{0.95,0.95,0.92}
\lstdefinestyle{mystyle}{
	backgroundcolor=\color{backcolour},   
	commentstyle=\color{codegreen},
	keywordstyle=\color{magenta},
	numberstyle=\tiny\color{codegray},
	stringstyle=\color{codepurple},
	basicstyle=\ttfamily\footnotesize,
	breakatwhitespace=false,         
	breaklines=true,                 
	captionpos=b,                    
	keepspaces=true,                 
	numbers=left,                    
	numbersep=5pt,                  
	showspaces=false,                
	showstringspaces=false,
	showtabs=false,                  
	tabsize=2
}
\usepackage{soul}
\definecolor{light-gray-x}{gray}{0.9}
\definecolor{pink-x}{rgb}{0.858, 0.188, 0.478}
\definecolor{blue-x}{RGB}{24, 90, 188}
\definecolor{gray-x}{RGB}{241, 243, 244}

\DeclareRobustCommand{\hllgrayx}[1]{{\sethlcolor{gray-x}\hl{#1}}}
\DeclareRobustCommand{\codex1}[1]{{\textbf{\textcolor{blue-x}{\small{\hllgrayx{#1}}}}}}

\newcommand{\mnflow}{\mbox{\codex1{\textit{mnFlow}}}}

\title{A Combinatorial Approach to Novel Boundary Design in Deterministic Lateral Displacement}

\author[$\dagger *$]{Aryan Mehboudi} 
\author[$\dagger$]{Shrawan Singhal} 
\author[$\dagger\ddagger$]{S.V. Sreenivasan}

\affil[$\dagger$]{NASCENT Engineering Research Center,
	The University of Texas at Austin, Austin, Texas 78758, United States}
\affil[$\ddagger$]{Walker Department of Mechanical Engineering, The University of Texas at Austin, Austin, TX 78712, United States}
\affil[$*$]{Email: aryan.mehboudi@austin.utexas.edu}

\date{March 31, 2025}

\begin{document}
	\clearpage
	\maketitle
	\thispagestyle{empty}
	
	\begin{abstract}
Deterministic lateral displacement (DLD) is a high-resolution separation technique used in various fields.
A fundamental challenge in DLD is ensuring uniform flow characteristics across channel, particularly near sidewalls where pillar matrix inevitably loses its lateral periodicity.
Despite attempts in the literature to improve boundary design, significant variations in critical diameter persist near sidewalls, adversely affecting the separation performance.
We propose a combinatorial framework to develop an optimal design aimed at minimizing flow disturbances.
We employ a set of parameterized boundary profiles, integrating multiple DLD channels, each with distinct design parameters, into a single microfluidic chip in parallel.
Fluorescent beads are introduced into the chip via through-wafer via, flowing through inlet buses and DLD channels.
The width of large-particle-laden stream downstream of channels is determined using fluorescence microscopy and image processing.
The experimental results suggest an optimal range of design parameters for depletion and accumulation sidewalls.
We conduct numerical simulations to further explore the experimental findings and refine the optimization.
Comparison of results with existing design methodologies in the literature demonstrates the superior performance of the proposed framework.
This work paves the way for design of DLD systems with enhanced performance, particularly for applications requiring high recovery rates and purity simultaneously.
\end{abstract}


	\newpage
	{
		\hypersetup{
			linkcolor=black, 
		}
		\tableofcontents
	}


\section{Introduction}
\label{sec_introduction}

\begin{figure*}[!bt]
	\centering
	\includegraphics[width=\textwidth]{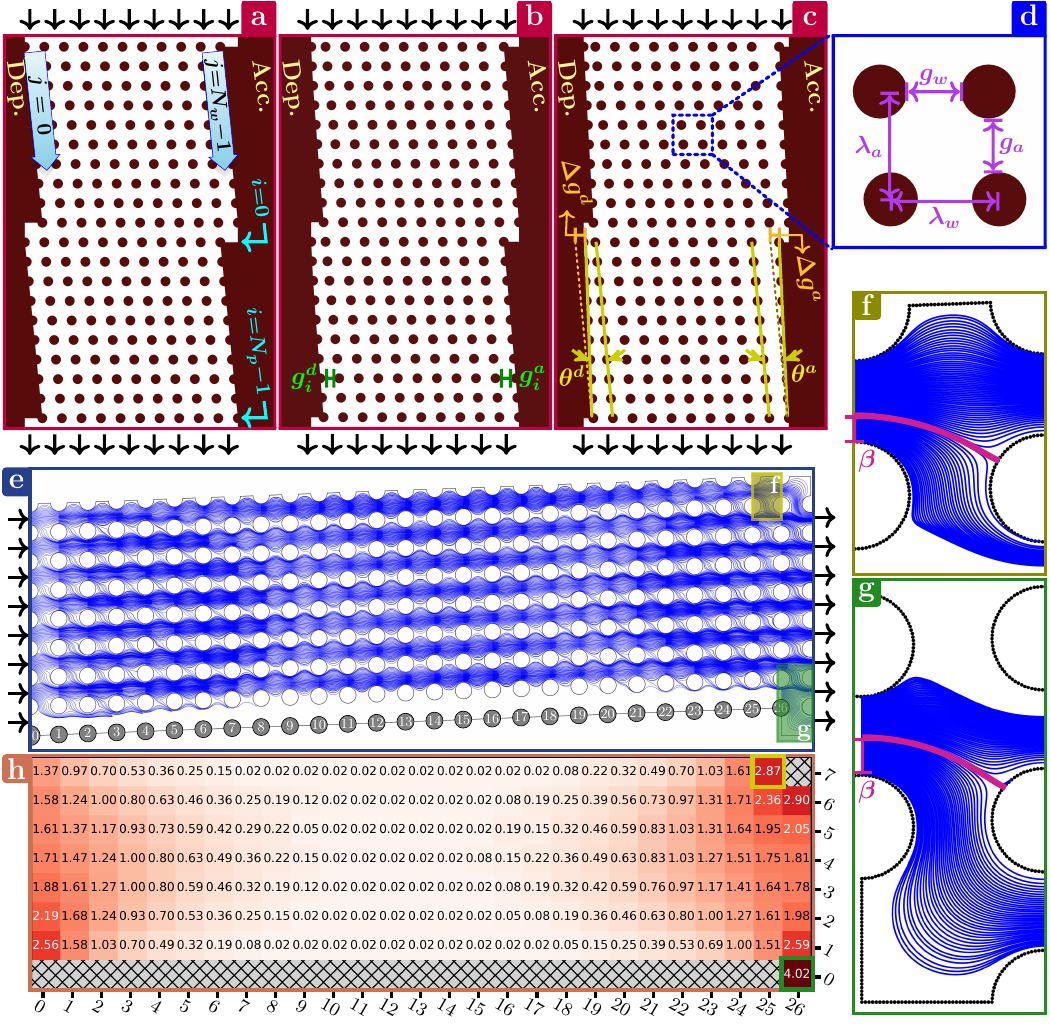}
	\caption{
		Geometrical configurations of DLD array and streamlines.
		Schematic representation of 
		a DLD structure with two full units, each consisting of $N_p$ rows of pillars and $N_w$ fluidic lanes, periodically repeating along the channel axis 
		without any special boundary treatment (a), 
		\textit{i.e.}, all boundary gaps equal that in the bulk of grid,
		together with 
		its counterpart  
		wherein
		depletion ($g^d_i$) and accumulation ($g^a_i$) boundary gaps on Row \#$i$ are determined from an analytical model from the literature~\cite{feng_maximizing_2017}~(b), 
		and the new class of parameterized boundary design proposed in this work (c), wherein 
		boundary gaps vary along the channel axis with adjustable parameters dictating
		the deviations of depletion ($\theta^d$ or $\Delta g^d$) and accumulation ($\theta^a$ or $\Delta g^a$) sidewalls from their unmodified positions (dotted lines).
		(d) Definition of geometric parameters for bulk of DLD array.
		(e) Fluid flow streamlines
		for different unit cells of a DLD system with
		$N_p{=}27$, $N_w{=}8$, $g_w{=}g_a{=}4.3~\mu m$, and $\lambda_w{=}\lambda_a{=}10~\mu m$ 
		together with the magnification of exemplary unit cells next to accumulation (f) and depletion (g) sidewalls, and 
		the corresponding local critical diameter 
		nondimensionalized by the value estimated from a Poiseuille flow-based model \cite{inglis_critical_2006}, \textit{i.e.}, ${\sim} 1.0~\mu m$, 
		showing large variations throughout the channel (h).
		The unit cells wherein the concepts of first stream width $\beta$ and local critical diameter are not applicable are hatched with lines in \textbf{h}. 
	}
	\label{fig_schem}
\end{figure*}

Deterministic lateral displacement (DLD) is a high-resolution separation technique developed by Huang \textit{et. al.} \cite{huang_continuous_2004}.
It has been successfully used for separation of 
DNA molecules \cite{wunsch_gel-on-a-chip_2019},
exosomes \cite{wunsch_nanoscale_2016, smith_integrated_2018},
droplets \cite{tottori_separation_2017},
blood components \cite{davis_deterministic_2006,zeming_label-free_2021}, 
red blood cells \cite{ranjan_dld_2014, zeming_asymmetrical_2016}, 
white blood cells \cite{liu_highly_2020,chavez-pineda_portable_2024}, 
circulating tumor cells \cite{xiang_precise_2019, liu_cascaded_2021}, 
etc.
A schematic representation of a basic DLD structure 
with $N_p$ rows of pillars and $N_w$ fluidic lanes 
per each full unit, 
periodically repeating along the channel axis is shown in \textbf{Figure~\ref{fig_schem}} (a\textendash d).
Each full DLD comprises a pillar matrix characterized by axial ($g_a$ and $\lambda_a$) and lateral ($g_w$ and $\lambda_w$) gap and pitch, respectively.
The pillar matrix is displaced laterally normal to channel axis.
This geometric configuration enables the separation of particles based on size: sufficiently large particles exhibit a ``bump'' motion, following the array tilt angle ($\tan^{-1} (\lambda_w/N_p\lambda_a)$), while smaller particles undergo a ``zigzag'' motion along the channel axis.

The emergence of these two distinct particle trajectories arises from the behavior of fluid flow through the DLD structure. As fluid flows through a gap between two adjacent pillars on the same row of a DLD structure with a periodicity of $N_p$, it splits into $N_p$ distinct streams, delineated by stagnation streamlines. For a particle with a radius larger than the width of the first stream adjacent to a pillar ($\beta$), its center lies outside the first fluid stream. Consequently, despite being near the pillar, the particle follows the second (or a subsequent) stream. This causes larger particles to theoretically avoid penetrating the lateral gaps between pillars. Instead, they move towards the accumulation sidewall in a bump mode trajectory, while being depleted near the depletion sidewall. 
The size threshold that defines these distinct behaviors is referred to as ``critical diameter'' ($d_c$), which depends on the periodicity of DLD array and other geometric parameters, such as pillar diameter and array pitch. For further details, readers are referred to comprehensive reviews, such as that by Salafi \textit{et. al.}~\cite{salafi_review_2019}.

One of the fundamental challenges in DLD arises from the sidewalls, which inevitably intersect with the pillar array and disrupt the lateral periodicity of the structure.
This disturbance in the fluid flow alters the local flow dynamics near the sidewalls.
Numerical simulations suggest that the sidewall effects diminish at approximately 20 unit cells away from the boundary~\cite{kim_broken_2017}.
The disturbance caused by sidewalls adversely affects the local first stream width and critical diameter characteristics.
The emergence of such nonuniformity in the local critical diameter reduces sorting efficiency, as particles of a given size are no longer directed uniformly in the same direction across the entire channel~\cite{pariset_anticipating_2017}.
This issue becomes particularly pronounced in DLD systems with a relatively small number of fluidic lanes, where the sidewalls can significantly influence a substantial portion of the channel. It can also present challenges in architectures where particles accumulate near the sidewalls, such as the condenser array introduced by Kim~\textit{et al.}~\cite{kim_broken_2017}.
%

Several studies in the literature have investigated the effects of sidewalls~\cite{inglis_efficient_2009,pariset_anticipating_2017,feng_maximizing_2017,ebadi_efficient_2019,inglis_fluidic_2020}.
Among these, Inglis~\cite{inglis_efficient_2009}, in one of the early studies, demonstrated that the absence of an appropriate boundary profile can significantly impair the ability of DLD to laterally displace large particles.
To elucidate this challenging phenomenon, we consider a DLD system with
$N_p=27$, $N_w=8$, $g_w=g_a=4.3~\mu m$, and $\lambda_w=\lambda_a=10~\mu m$.
By numerically solving the fluid flow equations and performing particle tracking simulations, we determine the variations in the first stream width ($\beta$) across the channel. This analysis provides deeper insight into how the critical diameter varies locally throughout the channel, as illustrated in Figure~\ref{fig_schem} (e\textendash h).
In the absence of an appropriate boundary design methodology, a significant portion of the channel in the middle of a full unit, \textit{i.e.}, row indices ranging approximately from 10 to 15, exhibits a very small critical diameter, which is advantageous for separation. However, the critical diameter increases substantially near the interfaces of full units and in proximity to the sidewalls. 
A large local critical diameter can hinder the displacement mode of relatively large particles that would otherwise migrate toward the accumulation sidewall, adversely affecting the separation performance of the DLD device.
Ideally, the maximum value of $d_c$ across each of the $N_w$ fluidic lanes should be minimized to an equally small value.
As an illustrative case, a particle must exceed $2.59~\mu m$ (the maximum $d_c$ in Lane \#1) to maintain its displacement mode while traversing Lane \#1.
Similarly, only particles larger than approximately
$4.02~\mu m$ (${\sim}93.5\%$
of the gap size) can escape Lane \#0 and transition to higher lanes via the bump mode.
Consequently, large-size constituents of the sample cannot be effectively displaced laterally, as a significant fraction remains within lanes exhibiting relatively high values of maximum-over-lane $d_c$.
This numerical example highlights the fundamental nature and critical significance of the problem addressed in this study.

A common approach to mitigating fluid flow disturbances near the sidewalls is to estimate appropriate boundary gap values using physics-based models, an example of which based on a model developed by Feng.~\textit{et. al.}~\cite{feng_maximizing_2017} is illustrated in Figure~\ref{fig_schem} (b). 
Inglis \cite{inglis_efficient_2009} demonstrated that proper configuration of sidewalls can reduce critical diameter variations and enhance separation performance.
An analytical model was developed to estimate appropriate boundary gaps, assuming that the fluid flux through a gap is proportional to the square of the gap size.
Thereafter, Feng.~\textit{et. al.}~\cite{feng_maximizing_2017} revised the model by assuming that the fluid flux through a gap is proportional to the cube of the gap size, analogous to fluid flow between shallow, infinitely wide channels.
A key limitation of these models is that plane Poiseuille flow may not accurately represent the fluid flow dynamics around the DLD pillars.
To address this, Ebadi~\textit{et. al.}~\cite{ebadi_efficient_2019} developed a correlation between gap size and fluid flux for a DLD unit cell by conducting two-dimensional numerical simulations for various gap sizes and performing curve fitting on the resulting dataset.
They established a correlation of the form $Q\propto g_w^b$, where $Q$ represents fluid flux, and $b$ is a fitting parameter reported to be approximately 2.46.
This value falls between the exponents used by Inglis \cite{inglis_efficient_2009} (2.0) and Feng.~\textit{et. al.}~\cite{feng_maximizing_2017} (3.0).
More recently, Inglis~\textit{et. al.}~\cite{inglis_fluidic_2020}
employed three-dimensional simulations of DLD unit cells with varying aspect ratios to derive a correlation between boundary gaps and the geometric configuration of the pillar matrix.

We have recently performed a numerical investigation of state-of-the-art boundary design methodologies~\cite{mehboudi_investigation_2025a}.
Although these models enhance the uniformity of fluid flow characteristics, significant nonuniformity can persist near the sidewalls.
While analytical models offer valuable insights into the physics of fluid flow, existing simplified analytical models appear unable to effectively resolve the disturbances induced by the sidewalls.
%
To the best of our knowledge, for the first time, we propose a novel paradigm in this work to address the aforementioned challenge.
Rather than relying on physics-based models, we adopt a set of parameterized profiles for boundary gap variations along the channel axis, as illustrated in Figure~\ref{fig_schem} (c), and conduct combinatorial studies to identify optimal design parameters.
We employ our recently developed DLD design automation (DDA) tool, a component of the micro/nanoflow (\mnflow) package \cite{mehboudi_universal_2025,mehboudi_mnflow_2024}, to generate a batch of diverse geometric configurations, which are subsequently integrated into the computer-aided design (CAD) mask layout of a microfluidic chip.
The microfluidic chip is fabricated using standard semiconductor fabrication processes. The performance of the DLD designs is experimentally characterized using fluorescent beads. Following the identification of the desired design parameter subspace, numerical simulations are conducted to further explore the experimental results and optimize the identified design parameter subspace.



\section{Results and Discussion}
\label{sec_res}

\subsection{Experimental Combinatorial Study}
\label{sec_exp_combinatorial}


\begin{table*}[!tb]
	\centering
	\caption{Geometrical configurations of various boundary design schemes explored experimentally in this work. Designs \#4 and higher are members of the design family proposed in this work as depicted in Figure~\ref{fig_schem} (c), wherein  
	$\Delta g_\textmd{Dep.}^\ast\equiv\Delta g^d/g_w$ and 
	$\Delta g_\textmd{Acc.}^\ast\equiv\Delta g^a/g_w$.
	Positive and negative values denote gap widening and shrinkage, respectively.
	}
	\begin{tabular*}{\textwidth}{@{\extracolsep{\fill}}||c c c c c c c ||} 
		\hline
		Design \# & $\Delta g_\text{Dep.}^\ast$ & $\Delta g_\text{Acc.}^\ast$ & & Design \# & $\Delta g_\text{Dep.}^\ast$ & $\Delta g_\text{Acc.}^\ast$ \\ [0.5ex] 
		\hline\hline
		1 & \multicolumn{2}{l}{No special boundary adjustments} & & 20 & -0.833 & 0.4 \\
		2 & \multicolumn{2}{l}{Gap-squared dependency is assumed for flux~\cite{inglis_efficient_2009}} & & 21 & -0.833 & 0.5 \\
		3 & \multicolumn{2}{l}{Gap-cubed dependency is assumed for flux~\cite{feng_maximizing_2017}} & & 22 & -0.833 & 0.6 \\
		4 & -0.667 & 0.252 & & 23 & -0.833 & 0.7 \\
		5 & -0.667 & 0.0 & & 24 & -0.833 & 0.8 \\
		6 & -0.667 & 0.1 & & 25 & -0.833 & 0.9 \\
		7 & -0.667 & 0.2 & & 26 & -0.833 & 1.0 \\
		8 & -0.667 & 0.3 & & 27 & -1.0 & 0.0 \\
		9 & -0.667 & 0.4 & & 28 & -1.0 & 0.1 \\
		10 & -0.667 & 0.5 & & 29 & -1.0 & 0.2 \\
		11 & -0.667 & 0.6 & & 30 & -1.0 & 0.3 \\
		12 & -0.667 & 0.7 & & 31 & -1.0 & 0.4 \\
		13 & -0.667 & 0.8 & & 32 & -1.0 & 0.5 \\
		14 & -0.667 & 0.9 & & 33 & -1.0 & 0.6 \\
		15 & -0.667 & 1.0 & & 34 & -1.0 & 0.7 \\
		16 & -0.833 & 0.0 & & 35 & -1.0 & 0.8 \\
		17 & -0.833 & 0.1 & & 36 & -1.0 & 0.9 \\
		18 & -0.833 & 0.2 & & 37 & -1.0 & 1.0 \\
		19 & -0.833 & 0.3 & &  &  &  \\
		[1ex] 
		\hline
	\end{tabular*}
	\label{tab_design_params}
\end{table*}

In this work, we consider configurations where the gaps between pillars are sufficiently smaller than the pillar height for the flow to be predominantly two-dimensional (2D). For 3D DLD structures, the fluid flow dynamics becomes independent of channel depth ($h$) when the normalized depth ($h^\ast\equiv h/\lambda_w$) exceeds $\sim$2.5~\cite{mehboudi_investigation_2025a}, a condition that is typically met to increase the throughput of DLD devices. Our fabricated channels meet this condition ($h^\ast\approx 8$), ensuring 2D-dominated behavior while enhancing throughput. Numerical simulations thus employ a 2D formulation for consistency.

We integrate all DLD structures with varying designs into a single microfluidic chip, significantly improving the efficiency of both fabrication and experimental examination compared to individually fabricating and operating each DLD system. To achieve this, we employ a parallelized architecture enabled by through-silicon via (TSV) technology, as detailed in Section~\ref{sec_exp_fab}.

Each microfluidic chip comprised 73 DLD channels, including duplicates and channels reserved for internal process development.
Herein, we report 37 distinct designs, the configurations of which are detailed in \textbf{Table~\ref{tab_design_params}}.
The batch included designs with no special boundary treatment ($None$; Design \# 1),
gaps estimated based on the gap-squared ($Pow_2$; Design \#2) \cite{inglis_efficient_2009},
and gap-cubed ($Pow_3$; Design \#3) \cite{feng_maximizing_2017} approximations, 
as well as various members of the parameterized design family illustrated in Figure~\ref{fig_schem} (c).

We define the dimensionless design parameters $\Delta g_\text{Dep.}^\ast\equiv\Delta g^d/g_w$ and $\Delta g_\text{Acc.}^\ast\equiv\Delta g^a/g_w$ 
at the upstreammost row ($i{=}0$) of the depletion and accumulation sidewalls, respectively. Here, negative and positive values denote the shrinkage and widening of the boundary gap compared to that in the bulk of the grid $g_w$, respectively.
Design \#4 (${\sim}Pow_3$) aims to produce linear boundary gap profiles, with boundary gaps at the upstreammost row of
the depletion ($\Delta g_\text{Dep.}^\ast{=}-0.667$) and
accumulation ($\Delta g_\text{Acc.}^\ast{=}0.252$) sidewalls of each full DLD unit matching those in the nonlinear $Pow_3$ design.
Additionally, there are three parameter sets with
$\Delta g_\text{Dep.}^\ast{=}-0.667$ (Designs \#5\textendash 15), 
$-0.833$ (Designs \#16\textendash 26), and 
$-1.0$ (Designs \#27\textendash 37). 
Each set comprises 11 designs with 
$\Delta g_\text{Acc.}^\ast$ ranging from 0 to 1 in increments of 0.1.

The design and manufacturing steps are configured such that the bulk of the DLD channels features 
$N_p{=}27$, $N_w{=}22$, $g_w{\approx}g_a{\approx}4.3\textendash4.6~\mu m$, and $\lambda_w{=}\lambda_a{=}10~\mu m$.
The critical diameter is estimated to be approximately $1.3~\mu m$ based on empirical data from the literature~\cite{davis_microfluidic_2008}.
%
We used samples consisting of PBS 1X with $2\%$ Tween 20, 
along with small (${\sim}1$\textendash$1.9~\mu m$) green and large (${\sim}1.7$\textendash$2.4~\mu m$) red particles for our experiments.
The Reynolds number is estimated to be $Re\approx 5.5\times10^{-3}$.

\begin{figure*}[!tb]
	\centering
	\includegraphics[width=\textwidth]{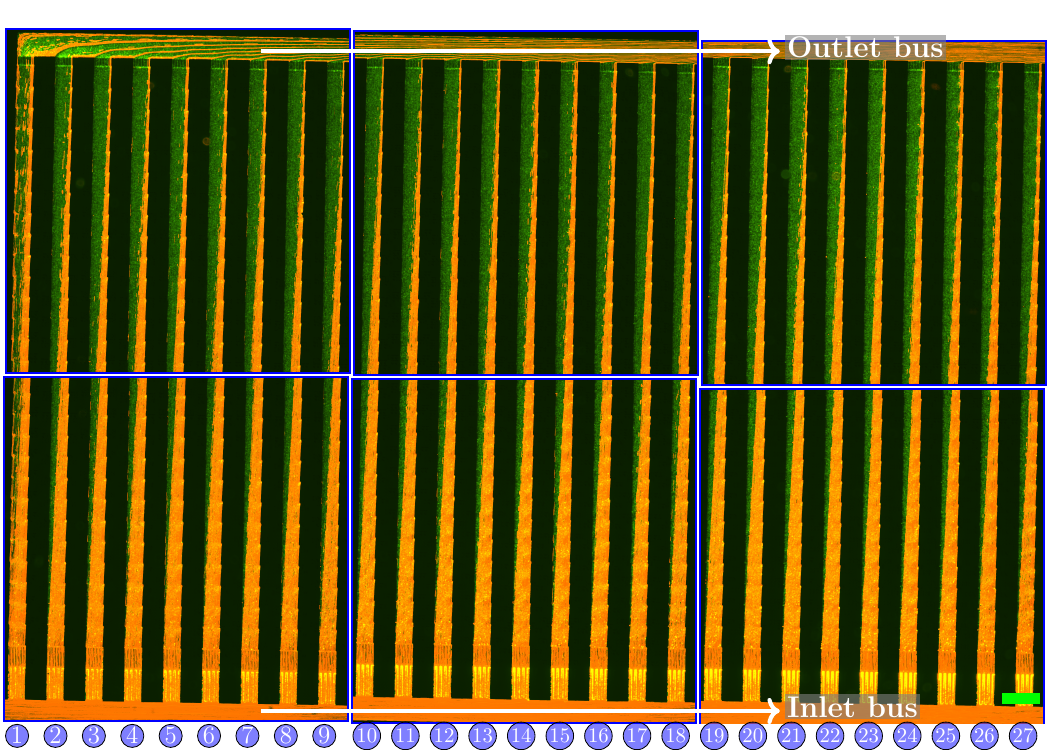}
	\caption{Stitched fluorescence microscopy images of the first 27 designs studied in this work experimentally, with their geometric configurations detailed in Table~\ref{tab_design_params}.
	Working fluid contains green and red polystyrene particles with size ranges of approximately $1.0~\mu m$\textendash$1.9~\mu m$ and 
	$1.7~\mu m$\textendash$2.4~\mu m$, respectively.
	Fluid flows from inlet bus (bottom) towards outlet bus (top) after passing through inlet filter, preload, and DLD arrays.
	Sufficiently large particles are laterally displaced towards accumulation sidewalls, while sufficiently small particles follow the average fluid flow direction.
	Design indices are labeled below the inlet bus for each channel. 
	Scale bar: $500~\mu m$.
	}
	\label{fig_res_microscopy}
\end{figure*}

Fluorescence microscopy images for the first 27 DLD channels are shown in \textbf{Figure~\ref{fig_res_microscopy}}. It can be observed that large red particles are concentrated near the accumulation sidewall of the DLD channels due to their bump-mode motion, in contrast to small green particles, which exhibit a zigzag-mode trajectory and create a uniform distribution throughout the channels.
It can also be observed that large particles adjacent to the depletion sidewall of Design \#1 are unable to transition to higher lanes and instead maintain a zigzag motion along the channel axis.
This observation aligns with the numerical simulation example presented in Figure~\ref{fig_schem} (e\textendash h) and underscores the necessity of proper DLD chip design near the sidewalls.

\begin{figure*}[!tb]
	\centering
	\includegraphics[width=.9\textwidth]{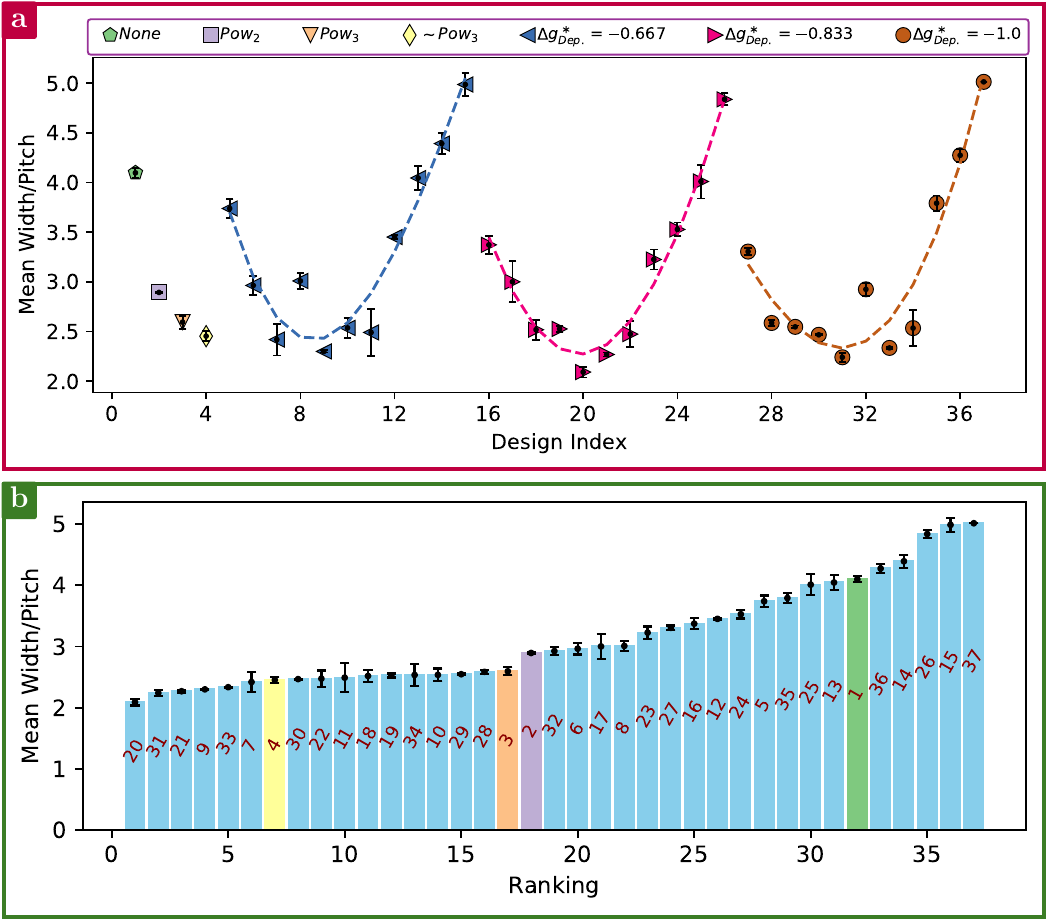}
	\caption{(a) Mean width of stream containing laterally-displaced particles concentrated next to accumulation sidewall downstream of DLD channels 
	nondimensionalized by pitch of DLD array ($w^\ast$), obtained experimentally through analyzing fluorescence microscopy images 
	related to downstream of each DLD system integrated into the fabricated microfluidic chip
	from multiple experiments 
	together with 
	(b) ranking of various designs based on their calculated performance metric $w^\ast$.
	Here, None, Pow\textsubscript{2}, Pow\textsubscript{3}, and  ${\sim}$Pow\textsubscript{3}, refer to Designs~\#1\textendash 4, respectively.
	Geometrical configurations of various designs are listed in Table~\ref{tab_design_params}.
	}
	\label{fig_res_comparison}
\end{figure*}

To evaluate the performance of various designs, we conducted image analysis on fluorescence microscopy images captured at relatively high magnification from the downstream region of the channels, as shown in the ESI. We calculated the width of the large-particle-laden (LPL) stream adjacent to the accumulation sidewall, averaging the results over the last approximately three full DLD units. The obtained results are presented in \textbf{Figure~\ref{fig_res_comparison}}.
It can be observed that a design parameter subspace near $\Delta g_\text{Acc.}^\ast{\approx}0.4$ results in a relatively narrower LPL stream downstream of the DLD channel, which can improve the separation performance in terms of purity and the concentration enhancement factor of large constituents.
Specifically, the average width of the LPL stream downstream of the channel decreases from approximately $28.9~\mu m$ and $25.9~\mu m$ for Designs \#2 and \#3, respectively, to ${\sim}20.9~\mu m$ for Design \#20, representing a reduction of approximately $19\%$\textendash$28\%$ in the width of the LPL stream.

\subsection{Numerical Investigation}
\label{sec_numerical_study}

Our goal in this section is twofold. First, we aim to investigate whether numerical modeling can corroborate the experimental observations. Second, we aim to explore the potential for further optimization of the parameterized design scheme developed in previous sections.
In most studies reported in the literature, the periodicity of DLD arrays falls within the range of 7 to 50. For optimization purposes, we consider a DLD structure with a moderate periodicity of $N_p{=}22$.
After identifying an optimal boundary design, we will apply it to DLD channels with a wide range of periodicity values to assess its applicability across different configurations.

\subsubsection{Optimization Refinement}
\begin{figure}
	\centering
	\includegraphics[width=0.7\textwidth]{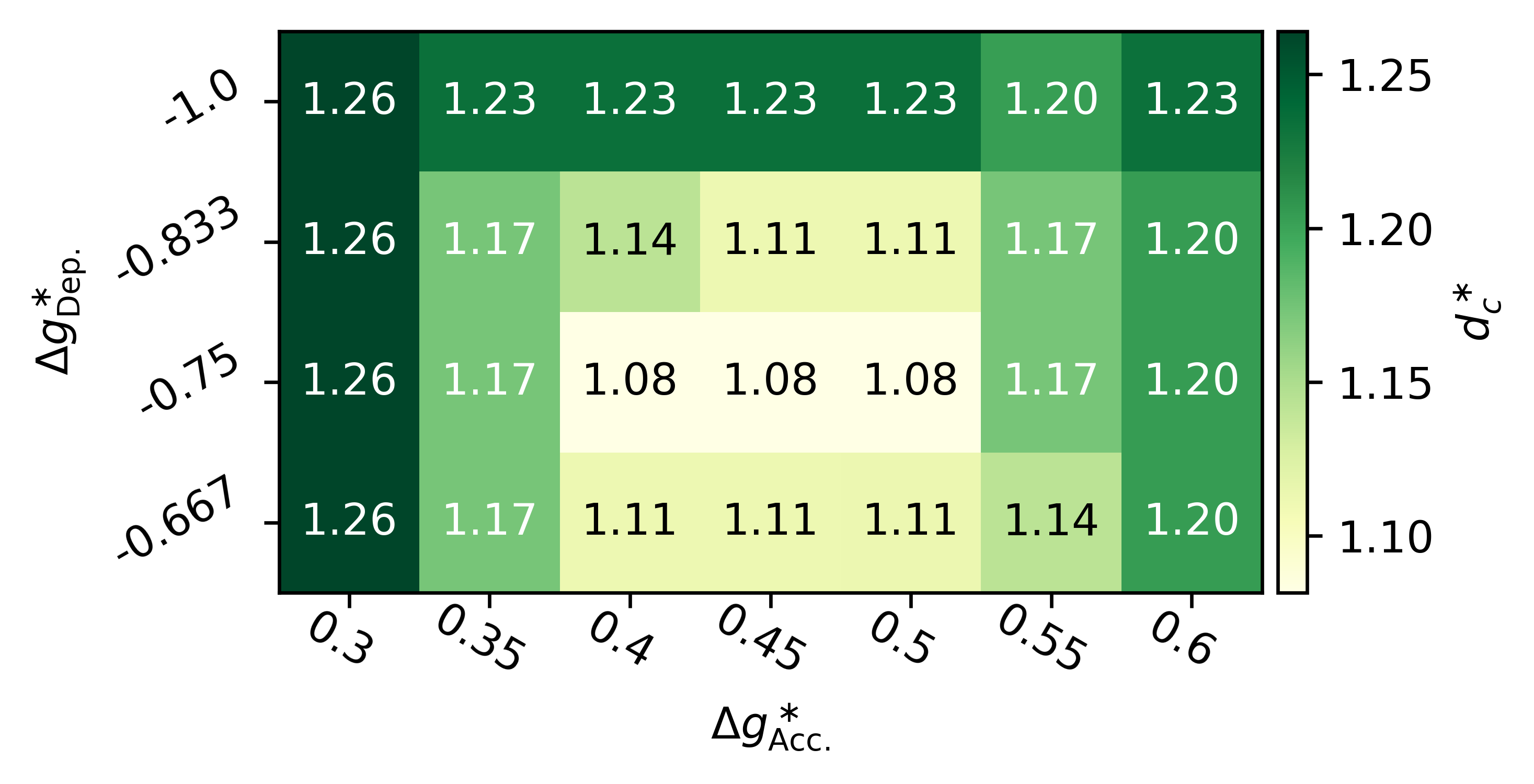}
	\caption{Maximum dimensionless critical diameter across entire DLD channel obtained through numerical simulations by using the parameterized boundary profile proposed in this work 
	with various depletion and accumulation design parameter values. 
	}
	\label{fig_res_heatmap}
\end{figure}

We conducted numerical simulations of DLD systems with
$N_p{=}22$, $N_w{=}8$, $g_w{=}g_a{=}7~\mu m$, and $\lambda_w{=}\lambda_a{=}14~\mu m$, 
incorporating various design parameters
$\Delta g_\text{Dep.}^\ast$ and 
$\Delta g_\text{Acc.}^\ast$ 
with a particular focus on the design parameter subspace identified by the experimental results.
We additionally employed a pressure balance scheme in proximity to the accumulation sidewall \cite{inglis_fluidic_2020}, as it has been demonstrated to further mitigate disturbances induced by the sidewalls, irrespective of the boundary gap distribution utilized~\cite{mehboudi_investigation_2025a}.
The maximum critical diameter found for each design point is 
nondimensionalized by an estimated value of  
ideal critical diameter
obtained from a Poiseuille flow-based model \cite{inglis_critical_2006} and is reported in \textbf{Figure~\ref{fig_res_heatmap}}.
It has been demonstrated that an optimal parameter subspace exists around
$\Delta g_\text{Dep.}^\ast{\approx}-0.75$ and $\Delta g_\text{Acc.}^\ast{\approx}0.45$,
minimizing the maximum local critical diameter across the channel.
While this observation aligns reasonably well with our experimental findings, these simulation studies allow for a more comprehensive optimization of the parameterized design scheme by analyzing parameter sets near the desired subspace, which was initially identified through experiments exploring a broader design parameter space with larger step sizes. 
Additionally, we found that a sidewall can significantly influence fluid flow characteristics at a considerable distance from the boundary. For relatively small $N_w$ values, the range of these effects can extend to the vicinity of the opposite sidewall, as demonstrated in the ESI.

\subsubsection{Generalization Evaluation}
\label{sec_res_comparison}
\begin{figure*}
	\centering
	\includegraphics[width=\textwidth]{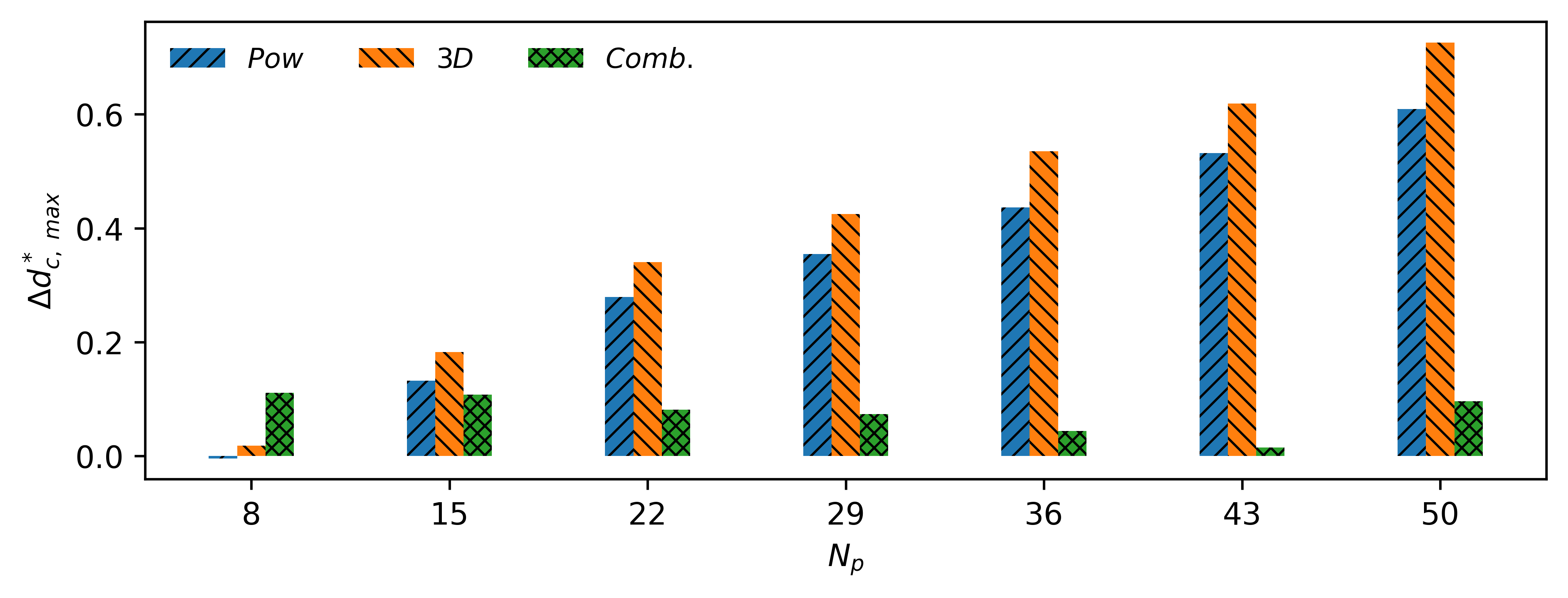}
	\caption{Variations of $\Delta d_{c,~max}^\ast$ with periodicity of DLD obtained from numerical simulations for the case that the boundary design is based on 
	the optimal design scheme determined in this work ($Comb.$), 
	the work of 
	Ebadi \textit{et. al.}~\cite{ebadi_efficient_2019} ($Pow$),
	and
	Inglis \textit{et. al.}~\cite{inglis_fluidic_2020} ($3D$).
	}
	\label{fig_res_comp_numerical}
\end{figure*}

We conducted simulations of DLD systems with varying periodicity levels, applying the optimal design scheme determined from the numerical combinatorial study, namely 
$\Delta g_\text{Dep.}^\ast{=}-0.75$ and 
$\Delta g_\text{Acc.}^\ast{=}0.45$.
For comparison, we also evaluated the most recent boundary design methods proposed by Ebadi \textit{et. al.}~\cite{ebadi_efficient_2019}
and
Inglis \textit{et. al.}~\cite{inglis_fluidic_2020}.
To quantify the deviation of the maximum critical diameter in a DLD channel ($d_c^\text{max}$) from an approximate ideal value ($d_c^\text{ref}$) based on a Poiseuille flow model~\cite{inglis_critical_2006}, we introduce the dimensionless metric $\Delta d_{c,~max}^\ast\equiv{(d_c^\text{max}-d_c^\text{ref})}/{d_c^\text{ref}}$. The results are presented in \textbf{Figure~\ref{fig_res_comp_numerical}}.
It can be observed that the optimal design scheme derived from the combinatorial study generalizes effectively across a wide range of DLD periodicity values. When employing the optimal boundary design, the deviation of the maximum critical diameter from the reference value, $d_c^\text{ref}$,
ranges between approximately $1\%$\textendash$11\%$, depending on the DLD periodicity.
Contours illustrating the corresponding variations in local critical diameter across the channels are provided in the ESI (Figs. S5--S11). 
While the $Pow$ and $3D$ schemes can achieve near-ideal fluid flow characteristics for sufficiently small DLD periodicities, such as $Np=8$, they can result in deviations as large as approximately $60\%$\textendash$70\%$ as the DLD periodicity increases towards $Np=50$.

\subsection{Implementation}
We have integrated the developed boundary design methodology into the DDA tool, enabling the design of a DLD channel with a single command, as demonstrated in the ESI through multiple examples. An illustrative code example for designing an entire DLD channel with a critical diameter of $d_c=10~\mu m$ and a periodicity of $N_p=40$ is presented in Listing~\ref{lst_code_mlb}.
The tool automatically configures certain parameters in the absence of explicitly provided arguments. Specifically, it assumes circular pillars with $g_w=g_a=d$, where $d$ represents the pillar diameter and is calculated to be approximately $42.0~\mu m$.
Additionally, the parameterized boundary design approach illustrated in Figure~\ref{fig_schem} (c) is a specific instance of the \textit{multi-lane boundary} (MLB) family, detailed in the ESI. This approach can be applied using the DDA tool to explore a broader design space in a combinatorial manner, potentially enabling the development of more effective designs as needed.

\noindent\begin{minipage}[!hb]{\linewidth}
	\begin{lstlisting}[language=Python, label={lst_code_mlb},
		caption={
			A single-command code example demonstrating the use of the DDA tool~\cite{mehboudi_mnflow_2024} to design a DLD system and generate its mask layout CAD file, where $d_c=10~\mu m$, $N_p=40$, and boundary gap profiles are determined using the optimal design scheme developed in this work through combinatorial analysis.
		}]
DLD(
	d_c=10, # Critical diameter (micron)
	Np=40,  # Periodicity
	boundary_treatment='mlb',
)\end{lstlisting}
\end{minipage}


\section{Conclusion}
\label{sec_conclusion}

In this work, we conducted experimental and numerical combinatorial investigations to develop a novel and effective boundary design approach aimed at tackling a critical issue in DLD systems\textemdash fluid flow disturbances near the sidewalls caused by the loss of lateral periodicity in the pillar matrix.
Rather than relying on simplified analytical models, we introduced a parameterized design family and explored its design parameters through combinatorial analysis. We prepared a batch of DLD designs with various combinations of these parameters and integrated all combinatorial designs into a single die in a parallel fashion. The microfluidic chips were fabricated using standard semiconductor processes, including photolithography and deep reactive ion etching on a silicon wafer, followed by sealing with a glass wafer via anodic bonding. We injected fluorescent beads into the device, conducted fluorescence microscopy, and performed image analysis to evaluate the performance of each design. Through this process, we identified a design parameter subspace with superior performance.
We performed numerical simulations which confirmed the experimental observations.
It also provided a further optimization of the parameterized design family with finer granularity.
Finally, we applied the developed design scheme to DLD systems with varying periodicity levels and compared its performance with state-of-the-art models from the literature. The results demonstrated the superior performance of our design scheme.

The optimal design scheme developed in this work minimizes the maximum deviation of the local critical diameter across the channel from an estimated ideal value to approximately $1\%\textendash 11\%$, 
depending on the DLD periodicity. 
While the current design demonstrates effectiveness, a more general variant of the parameterized boundary design methodology has been integrated into the DDA tool, enabling the configuration of multiple lanes near a sidewall by passing appropriate arguments in a single-command code as demonstrated in the ESI. 
Although the number of cases\textemdash and consequently the cost of pertinent computational and/or experimental combinatorial studies\textemdash can increase significantly with addition of design parameters, this framework can potentially be utilized to develop even more effective design techniques.
Furthermore, the gap size is presently assumed to vary linearly along the channel axis for each boundary lane adjacent to the sidewall, thereby reducing the costs associated with combinatorial studies. Nevertheless, the incorporation of additional design parameters for any boundary lane remains a potential avenue for future investigation.

We believe this work paves the way for the design of more effective DLD systems, particularly for narrow channels or applications where particles travel through the pillar array in close proximity to a channel sidewall, especially when high recovery rates and purity are simultaneously required.

\section{Experimental Section}
\label{sec_exp}


\subsection*{Design}
We designed various DLD systems using the DLD Design Automation (DDA) tool, a component of the Micro/Nanoflow ({\mnflow}) package~\cite{mehboudi_universal_2025,mehboudi_mnflow_2024}, and integrated them into a parallel fluidic circuit. As a safety measure, the DLD channels were elongated by three full units.

\subsection*{Fabrication of Microfluidic Chips}
\label{sec_exp_fab}
The chromium photomask with fine design features was acquired from Tekscend Photomask (formerly Toppan Photomasks), Inc., while the mylar masks for through-silicon vias (TSVs) were purchased from Fineline Imaging, Inc.
Photolithography and deep reactive ion etching (DRIE) were performed on the front side of the silicon wafer to fabricate fluidic circuits and DLD features with a depth of approximately $37~\mu m$. A similar process was then applied to the back side of the wafer to create inlet/outlet through-silicon vias (TSVs).
The processed silicon wafer was bonded to a glass wafer using anodic bonding, and the stack of bonded wafers was subsequently diced using a dicing saw.

\begin{figure*}
	\centering
	\includegraphics[width=\textwidth]{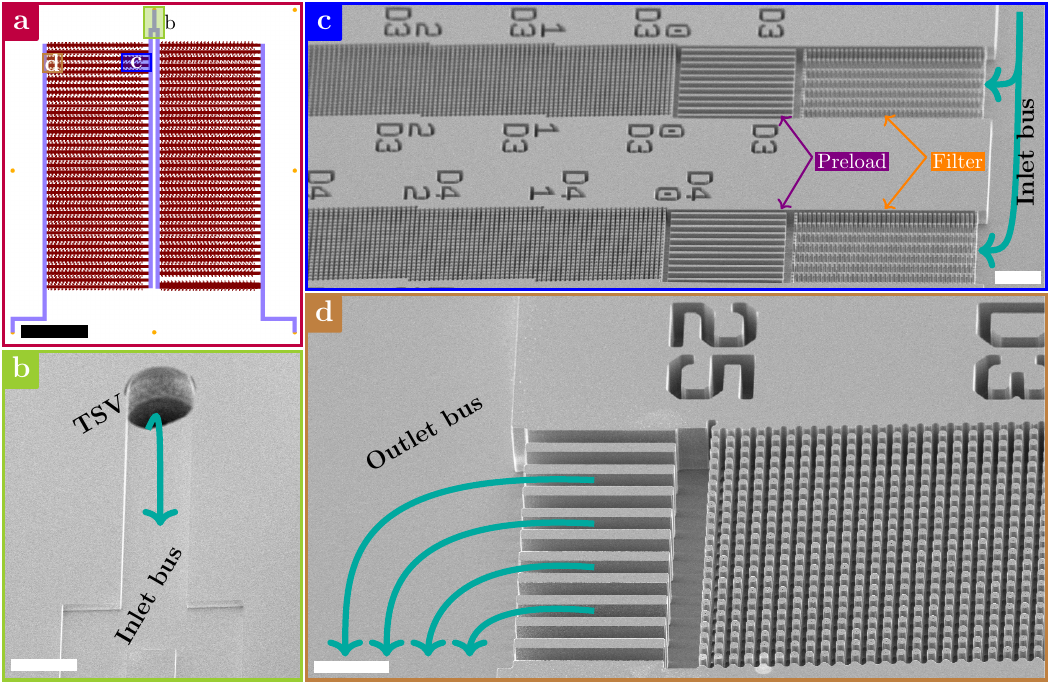}
	\caption{Schematic representation of computer-aided design (CAD) mask layout of device (a) together with SEM images of specified regions including 
		through-silicon via (TSV) and inlet bus (b), as well as
		upstream (c) and downstream (d) of channels.
		Scale bars are 5 mm, $300~\mu m$, $100~\mu m$, and $50~\mu m$ in a\textendash d, respectively.
	}
	\label{fig_fab}
\end{figure*}

A schematic representation of the computer-aided design (CAD) mask layout for the microfluidic chip, along with scanning electron microscopy (SEM) images of the processed wafer, is presented in 
\textbf{Fig.~\ref{fig_fab}}.
As shown in the figure, additional components are integrated upstream and downstream of the DLD structures. An inlet filter upstream of the DLD enhances chip longevity, while the preload component aligns particles and introduces them into the DLD region. Downstream, collection channels create a buffer zone between the DLD and outlet bus, with a feature size matching the DLD channel to ensure etch depth uniformity during fabrication. These channels also act as a coarse visual indicator, simplifying the identification of channels filled with the large-particle-laden stream downstream of the device.

\subsection*{Experiments}
Phosphate-buffered saline (PBS) (10X, Sigma-Aldrich) was diluted to 1X using deionized water. Device channels were primed with 1X PBS containing 2\% (w/v) Tween 20 (Sigma-Aldrich) before particle injection.
Green (1\% w/v, 1.0–1.9 $\mu m$) and red (1\% w/v, 1.7–2.4 $\mu m$) fluorescent microspheres (Spherotech) were diluted in 1X PBS with 2\% Tween 20 at a 1:100 ratio. A syringe pump controlled the flow rate during liquid injection, with a  volumetric flow rate of 20 $\mu$L/min unless otherwise specified.

\subsection*{Image Analysis}
The nd2 Python package \cite{lambert_tlambert03nd2_2025} was used to import Nikon fluorescence microscopy images into NumPy arrays, while the OpenCV package \cite{_opencvopencv-python_2025} was employed to detect contours of the large-particle-laden stream near the accumulation sidewall downstream of the channel, over a length approximately three times that of a full DLD unit. 
To ensure consistency and reduce potential bias, identical configurations were applied for all DLD designs.
The pixel counts of the contours were converted into the average stream width using the known pixel size of the high-magnification microscopy images ($\sim 0.55 \mu m$).

\subsection*{Numerical Simulations}

The DDA tool was used to prepare the CAD files for simulations. 
Ansys Fluent solved the 2D steady-state Navier-Stokes equations to obtain the pressure and velocity fields. Two full DLD units (each with $N_p$ rows) were modeled, with periodic boundary conditions at the inlet and outlet and no-slip conditions on all other boundaries.

Point-wise particle tracking was performed using a custom-developed code~\cite{mehboudi_tracking_2025a} to determine the following parameters:
first stream width ($\beta$), 
estimated critical diameter ($d_c\approx 2\beta$),
first stream flux fraction, 
and its dimensionless value 
normalized by its ideal counterpart ($1/N_p$).
The characteristics were evaluated across both full DLD units, showing consistent results. A mesh independence study for a reference case confirmed that the characteristics were mesh-independent.

Unless otherwise stated, the following configurations were used in the simulations.
The maximum local Reynolds number was approximately 0.03.
The tracer diameter was set to 1 nm.
The fluid viscosity was $1.003\times 10^{-3}$ Pa.s.
A total of 256 tracers were used per gap.
For the case of a $7~\mu m$-gap, therefore, the uncertainty in the calculated first stream width and local critical diameter was estimated to be approximately $\pm 13.7$ nm and $\pm 27.4$ nm, respectively.
%
%
%
We used a computer with 
16.0 GB RAM and a 
13th Gen Intel\textsuperscript\textregistered~Core\textsuperscript{TM} i9-13980HX Processor.
A typical fluid flow simulation took approximately 20 to 45 minutes to complete, depending on the case complexity and system workload.


	
	
\section*{Author Contributions}
\textbf{Aryan Mehboudi:} Conceptualization, Methodology, Software, Validation, Formal analysis, Investigation, Resources, Data Curation, Writing - Original Draft, Writing - Review \& Editing, Visualization, Project administration. 
\textbf{Shrawan Singhal:}  
Resources, 
Writing - Review \& Editing, 
Supervision. 
\textbf{S.V. Sreenivasan:}
Resources, 
Writing - Review \& Editing,
Supervision, Funding acquisition.

\section*{Conflicts of interest}
There are no conflicts to declare.
\section*{Data Availability Statement}
The developed code for particle tracking is accessible from 
\href{https://github.com/am-0code1/fspt}{https://github.com/am-0code1/fspt}.
%
The code for DDA is part of {\mnflow} package and can be found at \href{https://github.com/am-0code1/mnflow}{https://github.com/am-0code1/mnflow} with DOI:
\href{https://doi.org/10.5281/zenodo.14357811}{10.5281/zenodo.14357811}. 

\section*{Acknowledgements}

A.M. greatly acknowledges the Microelectronics Research Center (MRC) at The University of Texas at Austin and all technical staff for their contributions to the fabrication work. Fluorescence microscopy was performed at the Center for Biomedical Research Support Microscopy and Imaging Facility at UT Austin (RRID:SCR\_021756) by using a Nikon High Content Analysis fluorescence microscope system. A.M. greatly appreciates The CBRS Microscopy and Imaging Facility and all staff for their contribution to flow visualization work with special thanks to Anna Webb for training and assistance with the fluorescence microscopy.

	
	\providecommand{\href}[2]{#2}\begingroup\raggedright\endgroup



\end{document}